\documentclass[twocolumn,prd,nofootinbib]{revtex4-1}

\usepackage{latexsym,verbatim}
\usepackage{amssymb}
\usepackage{amsfonts}
\usepackage{amsmath,color}
\usepackage{subcaption}
\usepackage{graphicx}
\usepackage{bm}
\usepackage[T1]{fontenc}

\usepackage{hyperref}

\hypersetup{
    colorlinks=true,
    linkcolor=red,
    citecolor=red,
    filecolor=magenta,      
    urlcolor=cyan,
    pdftitle={Galactic Dynamics in General Relativity: the Role of Gravitomagnetism},
    pdfpagemode=FullScreen,}

\def\mb#1{\mathbf{#1}}

\def\ber{\begin{eqnarray}}
\def\eer{\end{eqnarray}}
\def\beq{\begin{equation}}
\def\eeq{\end{equation}}

\def\rmd{{\rm d}}

\def\ed{\end{document}}

\begin{document}

\author{Davide Astesiano}
\email{dastesiano@hi.is}
\affiliation{Science Institute, University of Iceland,
Dunhaga 3, 107 , Reykjav\'{\i}k, Iceland }

\author{Matteo Luca Ruggiero}
\email{matteoluca.ruggiero@unito.it}
\affiliation{Dipartimento di Matematica ``G.Peano'', Universit\`a degli studi di Torino, Via Carlo Alberto 10, 10123 Torino, Italy}
\affiliation{INFN - LNL , Viale dell'Universit\`a 2, 35020 Legnaro (PD), Italy}

\date{\today}

\title{Can General Relativity play a role in  galactic dynamics? }
\begin{abstract}
We use the gravitoelectromagnetic approach to the solutions of Einstein's equations in the weak-field and slow-motion approximation to investigate the impact of General Relativity on galactic dynamics. In particular, we focus on a class of the solutions for the gravitomagnetic field, and show that, contrary to what is expected, they may introduce non negligible corrections to the Newtonian velocity profile.
{These are the homogeneous solutions (HS) for the gravitomagnetic field, i.e. solutions with vanishing matter currents. We show how recent results about galactic dynamics are connected to this class of solutions.}
\end{abstract}

\maketitle

\section{Introduction}\label{sec:intro}

Galactic dynamics is normally studied using Newtonian gravity \cite{2008gady.book.....B} rather than General Relativity (GR), since for the largest part of its extension, far from the central zone, the gravitational field of a galaxy is expected to be weak  and the speed of the stars are small with respect to the speed of light. In this context, it is well known that the problem of the flatness of the rotation curves  \cite{rubin1978extended,sofue2001rotation} is one of the observations supporting the existence of Dark Matter (DM) \cite{strigari2013galactic,amendola2018cosmology}. In particular \cite{Ciotti:2022inn} DM is required at large galactocentric distances, as testified by radio observations in HI gas, while Newtonian models are in agreement with the observed rotation curves over a large fraction of the optical disk; in addition, DM is relevant for stability reasons \cite{ostriker}.

Nevertheless, it was recently suggested that GR effects might contribute to modify or totally eliminate the impact of DM in the description of rotation curves: in particular, we refer to the pioneering works by  \citet{Cooperstock:2006dt} and \citet{Balasin:2006cg}, where a solution of Einstein's equations was considered as a model of a galaxy.  The work by Cooperstock and Tieu  was criticised by \citet{Cross:2006rx,Menzies:2007dm}, while Balasin and Grumiller's solution was generalised by \citet{Astesiano:2021ren} and then used by \citet{crosta2020testing}, who obtained a good agreement with  GAIA data \cite{GAIA1,GAIA2} for the Milky way.  Post-Newtonian corrections in galactic dynamics were considered by \citet{Ramos-Caro:2012ren}, while  \citet{ludwig2021galactic} developed a model of galactic dynamics based on the equations that govern the motion of a weakly relativistic perfect fluid and used it to successfully reproduces galactic rotation curves without need for dark matter. 

Why should GR have an impact even though the system is not strictly ``relativistic''? The above papers share the underlying reason that GR adds a new degree of freedom, due to the role of mass currents, that has no Newtonian counterparts: this produces the so-called \textit{gravitomagnetic} effects (see e.g. \citet{ciufolini1995gravitation,Mashhoon:2001ir,Ruggiero:2002hz,Mashhoon:2003ax,Costa:2012cw}). The question is then to evaluate if these effects are large enough to actually change the dynamics:  according to a recent paper by  \citet{Ciotti:2022inn}, the answer is negative, since the gravitomagnetic  corrections within the matter distribution  are smaller by a factor $10^{-6}$ with respect to the usually adopted Newtonian profile. These conclusions are based on a very careful and detailed analysis, and do not depend on the specific model considered by the author, since their validity derives from the very relations between the gravitomagnetic field and its sources.  Ciotti's work is focused on the non-homogenous solutions (NHS) for the gravitomagnetic field, that are directly related to the sources by a Poisson equation: despite the interesting conclusions, the work misses a complete discussion about the homogenous solutions (HS), which naturally arise when we consider different limits of the general class of exact solution of Einstein's equations studied by \citet{Astesiano:2021ren}. Taking it seriously the HS, it is possible to show that they can provide contributions of the same order as the Newtonian ones, which we discussed in \citet{Astesiano:2022ozl} and called \textit{strong gravitomagnetism}.
{We show how the results of \citet{Cooperstock:2006dt}, \citet{Balasin:2006cg} and \citet{ludwig2021galactic} are due to the presence of the HS. Remarkably, the HS can also be of higher order in powers of $c^{-1}$ and therefore the gravitomagnetic limit can not be valid anymore. However these solutions are connected to the existence of singularities in the gravitomagnetic field.}

With the term HS (homogeneous solutions) we do not refer to a lack of spatial dependence of these solutions since, as we are going to show, they are indeed explicitly dependent  on both the  cylindrical coordinates $r$ and $z$; rather, in mathematical terms, we refer to the fact that they are solutions of the gravitomagnetic field equation with vanishing matter currents.

\section{The gravitomagnetic approach}\label{sec:limits}

The system that we are trying to describe must exhibit as much as possible stationarity and axial symmetry.
In the derivation of the equations of motion we assume that any internal symmetry-breaking structure, such as spirals or stellar bars, as well as any component with significant velocity dispersion, and hence pressure contribution, such as, e.g., stellar bulges, are neglected in favour of the analytic results. Moreover, we consider the velocity dispersion in the disc to be negligible compared to the main global rotational motion. Regarding stationarity, we note that inner bars can also buckle, leaving an echo in stellar motions of order a km/s for a Gyr or longer \cite{10.1093/mnras/stz3625}. Moreover, outer galaxies are warped, and impacts of dwarf galaxy further make the galaxy wobble \citep{10.1093/mnras/stac2571}. We assume this and other effects to be suppressed for now and we will consider the gas to be composed of baryonic matter only. The impact of our result is not affected by all the details that we need to add to create a more realistic model.\\
Therefore, we assume that the main dynamics of a galaxy with the mentioned properties  can  be modelled as a stationary and axially symmetric rotating dust fluid; this means that  we will consider o fluid of stars, 
which can themselves be treated as test masses and whose motion can be studied using the geodesic equation. Accordingly, as discussed by \citet{Ruggiero:2021lpf}, Einstein's field equations in weak-field and slow-motion approximation  can be written in the so-called gravitoelectromagnetic analogy in the form
\begin{eqnarray}
\nabla^{2} \Phi&=&-4\pi G\, \rho, \label{eq:poisson01} \\
\nabla^{2} \mb A & =& -\frac{8\pi G}{c}\, \mb j,  \label{eq:poisson02}  
\end{eqnarray}
where $\Phi$ is the gravitoelectric or Newtonian potential\footnote{Notice that $\Phi$ is defined in analogy with electromagnetism and differs by a minus sign from the standard definition of  the Newtonian potential.}, $\mb A$  the gravitomagnetic potential,  $\rho$ and $\mb j$ are the matter density and current, respectively. The gravitoelectromagnetic fields are usually defined as
\begin{eqnarray}
\mb E&=&-\bm \nabla \Phi, \label{eq:defE} \\
\mb B&=& \bm \nabla \times \mb A. \label{eq:defB}
\end{eqnarray}
In addition, the geodesic equation can be written in analogy with the Lorentz-like force equation
\beq
\frac{\rmd {\mathbf V}}{\rmd t}=-{\mathbf E}-2 \frac{{\mathbf V}}{c}\times {\mathbf B}, \label{eq:lor2}
\eeq
where $\mb V$ is the fluid element velocity.\footnote{Notice that all quantities are referred to an inertial system that is at rest with the center of mass of the galaxy.}  In summary, Eqs. (\ref{eq:poisson01}),(\ref{eq:poisson02}) and (\ref{eq:lor2}) can be used to describe the equilibrium of the dust fluid (remember that the mass current is defined by $\mb j = \rho \mb V$).  

Let us focus on Eq. (\ref{eq:poisson02}): its solution can be generally written in the form $\mb A=\mb A^{NHS}+\mb A^{HS}$ where $\mb A^{NHS}$ is the solution of the non-homogenous equation, while $\mb A^{HS}$ is the solution of the homogenous equation
\beq
\nabla^{2} \mb A  = 0. \label{eq:nablaHS}
\eeq
Accordingly, the gravitomagnetic field can be written in the form
\begin{equation}
    \mb B= \mb B^{NHS}+ \mb B^{HS}, \label{eq:defB1}
\end{equation}
where, in particular, $\mb B^{NHS}$ is strictly related to the sources described by $\mb j$. If we do not consider the presence of the additional component $\mb B^{HS}$, a self consistent analysis of the solution of the Eqs. (\ref{eq:poisson01}),(\ref{eq:poisson02}) and (\ref{eq:lor2}) within the mass distribution  leads to the results obtained by \citet{Ciotti:2022inn}, i.e. the smallness of the gravitomagnetic corrections with respect to the Newtonian behaviour. 

 On the contrary, here we are concerned with the impact of the homogenous-solution on the description of system dynamics. To this end, due to the symmetries of the system, we may set $\mb A^{HS}=A\mb e_{\phi}$, and the above equations (\ref{eq:poisson01}), (\ref{eq:nablaHS}) and (\ref{eq:lor2}) can be written in the adapted cylindrical coordinates $(r,z,\phi)$  in the form
\begin{eqnarray}
  \partial_{zz} \Phi+\partial_{rr} \Phi+ \frac{\partial_r \Phi}{r}&=& -4\pi G \rho, \label{GMPhi} \\
   \partial_{zz} \psi+ \partial_{rr} \psi- \frac{\partial_r \psi}{r}&=&0, \label{GMA0}\\
\frac{V}{r}\partial_z \psi&=& \partial_z\Phi \label{emz}, \\
    V^2&=&r(-\partial_r\Phi)+ V \partial_r\psi, \label{emr} 
  \end{eqnarray}
where we introduced the auxiliary function $\displaystyle \psi = 2r \frac{A^{}_{}}{c}$. In order to evaluate the impact of the HS, in what follows,  we solve the set of Eqs. (\ref{GMPhi})-(\ref{emr})  in the thin disk approximation. 

Before proceeding, we notice that the inclusion of pressure terms would not alter the main results: it would only make it harder to obtain analytic solutions. Despite the introduction of the pressure in many systems is necessary, they are not necessary to prove the non-negligibility of GR terms, which is the purpose of this paper. The equations, at the leading order in $c$, would be: 
\begin{align}
     \rho \left[-\frac{V}{r} \partial_z \psi+\partial_z \Phi\right]&=p_r, \label{SGMzzp}\\
     \rho \left[-\frac{V}{r} \partial_r \psi+\partial_r \Phi+  \frac{V^2}{r}\right]&=p_z\label{SGMrrp}\\
   \partial_{rr} \psi+ \partial_{zz} \psi- \frac{\partial_r \psi}{r}&=0, \label{Sgmsource1p}\\
    -\frac{1}{4\pi G} \nabla^2 \Phi&=\rho\label{Sgmsource2p}.
\end{align}
As we see, the homogeneous equation (\ref{Sgmsource1p}) for $\psi$  is unaltered, and the relevant  terms impact on the system dynamics in the same way and with the same interpretation.

\section{The homogeneous solution for a thin disk}\label{sec:hs}

To begin with, let us consider the equation for the Newtonian potential (\ref{GMPhi}). As it is well known \cite{2008gady.book.....B}, for a thin disk, we have
\begin{align}
    \Phi(r,z)= \int^\infty_0 \tilde{\Phi}(u) e^{-u|z|} J_0(ru) du,
\end{align}
where $J_m(ru)$ is the Bessel function of the order $m$ and we imposed the asymptotic boundary condition $\displaystyle \lim_{z\rightarrow\infty} \Phi=0$. The spectral density  $\tilde{\Phi}(u)$ is fixed by the other boundary condition for the potential:
\begin{align}
   \frac{1}{2\pi G} \sigma(r)=- \partial_z\Phi(r,0)=  \int^\infty_0 \tilde{\Phi}(u) u  J_0(ru) du, \label{sigma}
\end{align}
where $\sigma(r)$ is  the (superficial) density profile of the matter distrivution; we remark that this is just the application of the Gauss theorem. Using the inverse Hankel transform we can find $\tilde{\Phi}(u)$
\begin{align}
\tilde{\Phi}(u)= 2\pi G \int^\infty_0 r \sigma(r) J_0(ru) dr,  
\end{align}
which completely fixes the value of the Newtonian potential everywhere. \\

We now consider  the equations for the gravitomagnetic function $\psi$. In the same way as we have done for the Newtonian potential, the solution to the homogenous equation (\ref{GMA0}) is\footnote{There are alternatives solutions given in terms of modified Bessel function.}
\begin{align}
    \psi(r,z)=\int^\infty_0 r e^{-\lambda |z|}\tilde{\psi}(\lambda) J_1(\lambda r) d\lambda, \label{psi}
\end{align}
that holds for a generic spectrum function $\tilde{\psi}(\lambda)$. From a physical perspective, if we fix the behaviour of $\psi$ along one axis, we completely fix the function $\psi$ everywhere. This suggests to use the equation of motion along the $z$-axis (\ref{emz}) in the plane $z=0$ as a boundary condition for $\psi$. In this way we can find $\psi(\lambda)$  and completely fix the gravitomagnetic function $\psi$. We evaluate Eq. (\ref{emz}) on the equatorial plane $z=0$
\begin{align}
    \partial_z \psi(r,0)= \frac{\partial_z \Phi(r,0)}{V(r)} r.
\end{align}
Using the form (\ref{psi}) and, again, the Hankel transform, we obtain the result
\begin{widetext}
\begin{align}
    -\int^\infty_0 r\lambda  \tilde\psi(\lambda)  J_1(\lambda r) d\lambda= \frac{\partial_z \Phi(r,0)}{V(r)} r \implies \tilde{\psi}(\lambda)=- \int^\infty_0 r \frac{\partial_z\Phi(r,0)}{V(r)} J_1(\lambda r) dr. 
\end{align}
\end{widetext}
Remembering the relation between the derivative along $z$ of the Newtonian potential and the density profile given in Eq.(\ref{sigma}) we can rewrite the spectral density as
\begin{align}
    \tilde{\psi}(\lambda)=2 \pi G \int^\infty_0 r \frac{\sigma(r)}{V(r)} J_1(\lambda r) dr. \label{tildepsi}
\end{align}
We now have the machinery to find the Newtonian potential $\Phi$ and the gravitomagnetic function $\psi$ for a given set of density profile $\sigma$ and velocity profile $V$. The last equation that we have to solve, Eq. (\ref{emr}),  will give us the relation between the velocity profile $V$ and the density profile $\sigma$. From this equation we find
\begin{align}
    \int^\infty_0  r \lambda\, \tilde{\psi}(\lambda)\, J_0(\lambda r) d\lambda= V(r)+ \frac{r}{V(r)} \partial_r\Phi(r,0).
\end{align}
Using another Hankel transformation and the form of the spectral density $\tilde{\psi}$ in Eq. (\ref{tildepsi}), we eventually get
\begin{widetext}
\begin{align}
  2 \pi G \int^\infty_0 r \frac{\sigma(r)}{V(r)} J_1(\lambda r) dr= \int^\infty_0 \left[V(r)+ \frac{r}{V(r)} \partial_r \Phi(r,0)\right] J_0(\lambda r) dr. \label{Fe}
\end{align}
\end{widetext}
A possible solution is given equating the integrands
\begin{align}
    V(r)= \sqrt{- r \partial_r \Phi(r,0)+ 2\pi G r \sigma(r) \frac{J_1(\lambda r)}{J_0(\lambda r)}}. \label{v}
\end{align}
The solution represents an infinitely thin disc.
The deviation from Newtonian dynamic is given by the second term. We see, in fact, that the Newtonian relation for circular orbits $V^2/r= -\partial_r \Phi$ is restored for $\lambda=0$. The second term is well behaved for $r \rightarrow 0$ if the density does not go to infinity at the origin.\\

The above solution is unique, once that the available observables, the density and velocity profiles, are used in the equations. Moreover, the final expression for the velocity profile $V$ is directly given in terms of the density profile and the Newtonian potential: accordingly, there is no need to explicitly calculate the gravitomagnetic potential $A$ or $\psi$. We remark that we gave the simplest solution for $V$, despite a deeper analysis for different explicit cases can produce more solutions; in addition, there are some singularities for this solution, but they can be located far from the regions we are interested  in  and, then, the solution can be truncated at some distance $\bar r$. In general, to obtain a GR solution for any truly isolated and finite system, we always need to truncate the solution at some distance. Subsequently, the full set of
matching conditions between the interior metric and the exterior vacuum solution
must be imposed in order to analyse the physical problem. The construction of the global solution requires three steps:
namely, obtaining the interior metric, solving the exterior vacuum field
equations and imposing the matching conditions between the two spacetimes. The matching of the interior and exterior solutions of isolated rotating objects in equilibrium is a fundamental problem in GR. In general this is extremely complicated and unfortunately, despite many efforts, there is not a simple picture yet. In this case if we want to look at the exterior solution we should solve
Eqs. (\ref{GMPhi}) and (\ref{GMA0}) with $\rho=0$. In particular, the prescription to do the matching is given in the paper by \citet{Mars:1998qd}.

However different solutions of Eq. (\ref{Fe}) do not need to have these singularities. 

The direct solution in Eq.  (\ref{v}) is interesting also in the region where $ \lambda r \ll 1$. In this region the velocity is
\begin{align}
    V(r)= \sqrt{- r \partial_r \Phi(r,0)+\lambda \pi G  r^2 \sigma(r)}, \label{ES}
\end{align}
which is well-behaved.\\

\begin{figure}[h!]
  \centering
  \begin{subfigure}[b]{0.78\linewidth}
    \includegraphics[width=\linewidth]{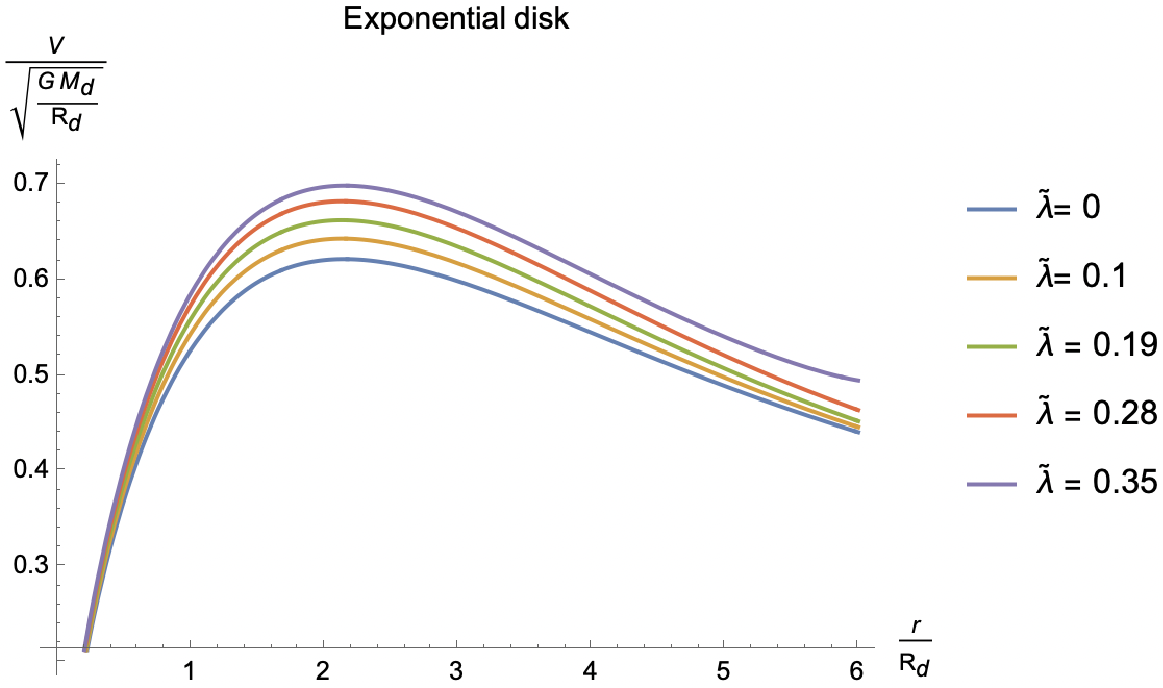}
    \caption{Exponential disk.}
  \end{subfigure}
  \begin{subfigure}[b]{0.78\linewidth}
    \includegraphics[width=\linewidth]{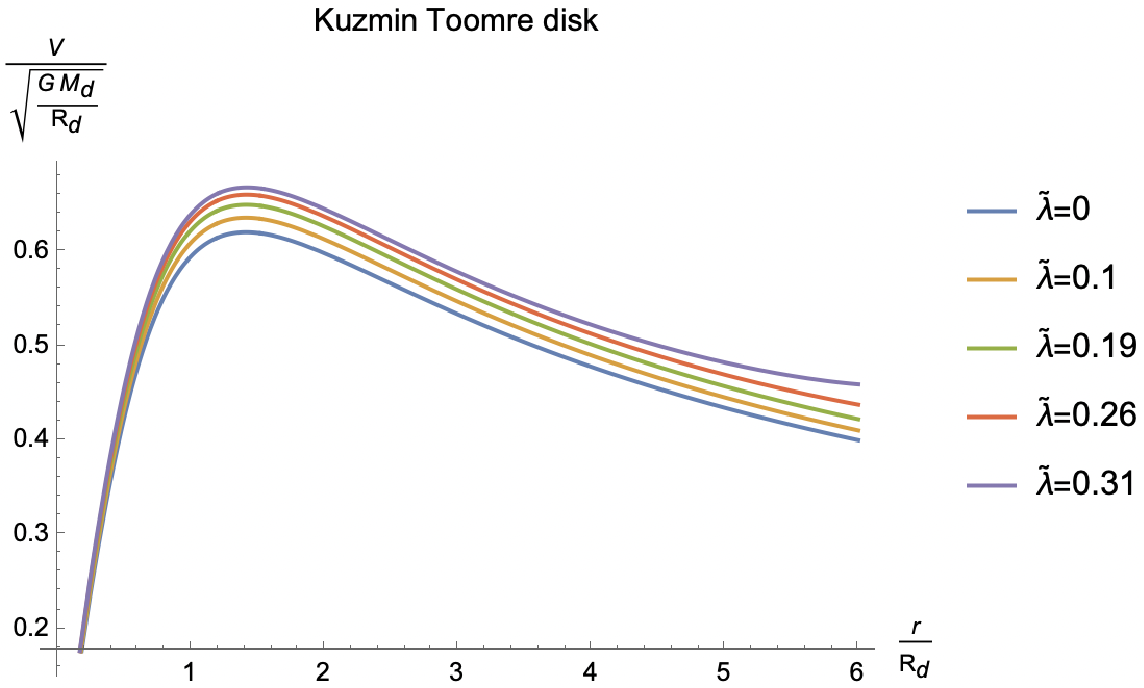}
    \caption{Kuzmin-Toomre disk.}
  \end{subfigure}
  \caption{The comparison between the Newtonian velocity profile in blue $(\tilde{\lambda}=0)$ and various GEM profile for different values of $\tilde{\lambda}$.}
  \label{fig:coffee}
\end{figure}

As we already stressed, these HS exist also in the general solution, where they affect the velocity even more than in the gravitomagnetic limit \cite{Astesiano:2022ozl}. One may wonder about the interpretation of these new solutions: the analogy with electromagnetism comes in handy. {If we impose that the system is stationary and asymptotically flat, the homogeneous solutions may be related to sources outside the region or due to the presence of singularities in the field $A$. For example in the case we are considering here, the singularities may come from the thin-disc approximation. }

When we write down the stationary equations, the history of the galaxy is stored in the boundary conditions that we impose for $A$. For example, the Newtonian case is recovered imposing the low energy limit plus the boundary condition $A=0$. For the same distribution of matter, we can have different solutions, depending on the boundary conditions for $A$. The solution to the Einstein's equations is unique once that we have fixed
these boundary conditions. In some sense, the gravitational field is ``carrying'' energy and momentum on its own, and this is also confirmed by the generalization of the virial theorem found in \citet{Astesiano:2022gph}:
\begin{align}
    \langle 2\int \rho V^2 d^3x-\frac{1}{8\pi G} \int \left(E^2+B^2\right) d^3x \rangle=0.
\end{align}
Using the analogy with electromagnetism we see that the second term is the total energy stored in the gravity fields. Therefore we have the balance equation
\begin{align}
    2 \times \text{energy of free dust (kinetic energy)} = \text{energy of gravity}.
\end{align}
This proposed interpretation for the HS requires further investigation. We plan to focus on this aspect and try to confirm these intuitions in forthcoming works.

\subsection{Application to the Kuzmin-Toomre model and the exponential disk}\label{ssec:apps}

The Kuzmin-Toomre model is a thin disk model, whose potential and surface density for a disk of total mass $M_d$ and scale length $R_d$ are given by \cite{2008gady.book.....B}
\begin{align}
    \Phi(r,0)=  \frac{G M_d}{\sqrt{(r^2+R_d^2)}}, \quad \sigma(r)= \frac{M_d R_d}{2 \pi (r^2+R_d^2)^{3/2}}.
\end{align}
The expression for the velocity obtained using Eq. (\ref{v}) turns out to be
\begin{align}
    V^{KT}(r)= \sqrt{\frac{G M_d}{R_d}} \sqrt{\frac{ \tilde{r} \left(\tilde{r} J_0(\tilde{r} \tilde{\lambda} )+ J_1(\tilde{r} \tilde{\lambda}
   )\right)}{\left(\tilde{r}^2+1\right)^{3/2} J_0(\tilde{r} \tilde{\lambda} )}},
\end{align}
where $\tilde r=r/R_d$ and $\tilde{\lambda}= R_d\,  \lambda$ are dimensionless parameters.\\ 

The exponential disk has the following potential-density pair
\begin{widetext}
\begin{align}
    \Phi(r,0)=\frac{G M_d\, r \left[I_0\left(\frac{r}{2 R_d}\right) K_1\left(\frac{r}{2 R_d}\right)-I_1\left(\frac{r}{2 R_d}\right) K_0\left(\frac{r}{2 R_d}\right)\right]}{2
   R_d^2}, \quad \sigma(r)= \frac{M_d}{2\pi R_d^2} e^{-r/R_d}.
\end{align}

where $K_{0},K_{1},I_{0}, I_{1}$ are modified Bessel functions \cite{2008gady.book.....B}. Using Eq. (\ref{v}) we obtain the following expression for the velocity:
\begin{align}
   V^{E}(r)= \sqrt{\frac{G M_d}{R_d}} \frac{\sqrt{\tilde{r} \left(\frac{2 e^{-\tilde{r}} J_1(\tilde{r} \tilde\lambda )}{J_0(\tilde{r} \tilde\lambda )}+\tilde{r} I_0\left(\frac{\tilde{r}}{2}\right)
   K_0\left(\frac{\tilde{r}}{2}\right)-\tilde r I_1\left(\frac{\tilde{r}}{2}\right)
   K_1\left(\frac{\tilde{r}}{2}\right)\right)}}{\sqrt{2}},
\end{align}
\end{widetext}
where again $\tilde r=r/R_d$ and $\tilde{\lambda}= R_d\,  \lambda$. 
The velocity profiles for the  Kuzmin-Toomre model and the exponential disk are in Figure \ref{fig:coffee}. In both cases, we see that the corrections to the Newtonian velocity profile (the curve in the bottom in both plots) are not negligible,  unlike what happens for the NHS, where they are in the order of 1 part to $10^{6}$  \cite{Ciotti:2022inn}.


\section{Relation with other works}
{The discussion above is restricted to the case of weak-field and slow-motion approximation of GR, described by the gravitoelectromagnetic limit. However, the HS are more general solutions to the equations of motion. In fact, we can consider a stronger regime for the HS which we called \textit{strong gravitomagnetism} and, as shown in \citet{Astesiano:2022ozl}, these geometric effects affect the dynamics and energy density even more. In this case, the analogue of the system of Eqs. (\ref{GMPhi})-(\ref{emr}) is:
\begin{widetext}
\begin{eqnarray}
   \left[\nabla^2 \Phi+ \frac{(\partial_{z}\psi)^2+(\partial_r\psi-2 \frac{\psi}{r})^2}{2 r^2}\right]&=&-4\pi G\rho\\
   \partial_{rr} \psi+ \partial_{zz} \psi- \frac{\partial_r \psi}{r}&=&0,\\
     \frac{V}{r} \partial_z \psi&=&\partial_z \Phi+ \frac{\psi}{r^2} \partial_z \psi, \label{SGMzz}\\
     \frac{V}{r} \partial_r \psi&=&\partial_r \Phi+  \frac{V^2}{r}+ \frac{\psi}{r} \partial_r \frac{\psi}{r}\label{SGMrr}
   \end{eqnarray}
\end{widetext}


The gravitomagnetic limit is restored if we consider a linear approximation with respect to $\psi$. 
In this limit of the general equations, the HS for $\psi$ have the same form as for standard gravitomagnetism, but
they affect even more the density and velocity. \\
The strong gravitomagnetic limit is related to the exact solutions used by  \citet{Cooperstock:2006dt} and \citet{Balasin:2006cg}; in order to understand this fact, we need to take
\begin{align}
    \Phi(r,z)= -\frac{1}{2}(r\Omega_0 -\frac{\psi(r,z)}{r})^2,\quad \Omega(r,z)=\Omega_0.
\end{align}
where $\Omega$ is the rotation rate of the dust.
In this way we get a rigidly rotating model. Eq.(\ref{SGMzz}) and (\ref{SGMrr}) are solved, while the others equations boil down to 
\begin{align}
    \frac{(\partial_{z}\psi)^2+(\partial_r\psi-2 r \Omega_0)^2}{r^2} =8 \pi G\rho, \\ 
    \partial_{rr} \psi+ \partial_{zz} \psi- \frac{\partial_r \psi}{r}&=0,\\
    \quad V&= r \Omega_0,
 \end{align}
which is indeed the low energy limit of the subclass used by \citet{Balasin:2006cg}, \citet{Cooperstock:2006dt} and  \citet{crosta2020testing},

In particular, the equations are exactly the same.\footnote{The only equation that is missing is a line integral which in the general system of equations fixes the coordinates $(r,z)$. Here in the low energy limit at leading order in $c^{-1}$ it becomes trivial. However, in the low energy limit this does not bring relevant corrections to the dynamics.} In these works $\Omega_0=0$,  which is equivalent to a rigid rotation of the reference frame and $\psi(r,z):=N(r,z)=v(r,z)\, r$ in their notation. {The variable $v(r,z)$, as discussed by \citet{Astesiano:2022gph}, is the velocity respect to the zero angular momentum observers (ZAMO) and, for this velocity, $v/r$ is not constant.}

\section{Comments and final remarks}\label{sec:concs}

In this paper we addressed the issue of the relevance of GR corrections to the galactic rotation curves; in the weak-field and slow-motion approximation which leads to the gravitoelectromagnetic analogy, we pointed out that there are solutions for the gravitomagnetic field that might have a relevant impact on the velocity profile. 
There is no contradiction with what was recently obtained by \citet{Ciotti:2022inn}, since Ciotti's results refer to the \textit{non-homogenous solutions} (NHS) for the gravitomagnetic potential, that are directly related to its sources, the mass currents, while our statement refers to the \textit{homogenous solutions} (HS), whose relation with the sources is determined by the equilibrium conditions of the dust fluid.  The origin of these HS is strictly related to the general solution of the Einstein's equations for a neutral, stationary and axisymmetric rotating dust \cite{Astesiano:2022ozl} and they are present also in the so-called strong gravitomagnetic regime. 
{We showed how the results of \citet{Cooperstock:2006dt}, \citet{Balasin:2006cg} and \citet{ludwig2021galactic} are due to the presence of the HS. However these solutions are connected to the existence of singularities in the gravitomagnetic field or due to de-localized sources.}

We showed that if the density profile is known, the equations can be analytically solved to obtain the velocity profile which  includes the GR corrections. As examples of  application of our approach, we explicitly calculated the velocity profiles for the  Kuzmin-Toomre model and the exponential disk, and showed that the impact of the gravitomagnetic corrections can be in the order of the 10\%-15\% and, hence, much greater than what is obtained for NHS case. From  theoretical considerations, here we noticed that $R_d \lambda<0.5$ for these models to avoid singularities in the galaxy model. An analysis of the data for different galaxies can help to better constrain the value of $\lambda$. It is important to remark that, from the perspective of the class of exact solutions, there could be limits in which the effect of GR is much greater than the one found here, for example in the above mentioned strong gravitomagnetic limit \cite{Astesiano:2022ozl}.

Our toy model shows how GR effects are not negligible in general. However, there is still work left to build realistic models. Since the existence of the homogeneous solutions in stationary systems is due to the presence of singularities, in order to build realistic models, a full analysis and study of which singularities we can consider physically admissible is needed. {Of course, major improvements are needed to understand the reliability of these solutions.} One of these further steps is to add pressure, in order to model a gas with a non negligible velocity dispersion in the stars (and gas) of the disc. Another interesting aspect would be to consider a dynamically hot stellar component modelling the bulges often present in the few central kpcs of massive disc galaxies.

In conclusion, we believe that our approach emphasises the impact of GR on galactic dynamics, determined by the rotational degree of freedom which is not present in Newtonian dynamics; the study of these effects may be relevant to better understand the dynamics of such systems and, consequently, the role played by dark matter.\\

\section*{Acknowledgments}
The work of D.A. is supported in part by Icelandic Research Fund grant 228952-051. The authors thank Luca Ciotti, L\'{a}rus Thorlacius, Valentina Giangreco Puletti, Fri{\dh}rik Gautason, Antonello Ortolan and Clive C. Speake  for stimulating discussions.

\bibliography{refs}

\begin{thebibliography}{28}%
\makeatletter
\providecommand \@ifxundefined [1]{%
 \@ifx{#1\undefined}
}%
\providecommand \@ifnum [1]{%
 \ifnum #1\expandafter \@firstoftwo
 \else \expandafter \@secondoftwo
 \fi
}%
\providecommand \@ifx [1]{%
 \ifx #1\expandafter \@firstoftwo
 \else \expandafter \@secondoftwo
 \fi
}%
\providecommand \natexlab [1]{#1}%
\providecommand \enquote  [1]{``#1''}%
\providecommand \bibnamefont  [1]{#1}%
\providecommand \bibfnamefont [1]{#1}%
\providecommand \citenamefont [1]{#1}%
\providecommand \href@noop [0]{\@secondoftwo}%
\providecommand \href [0]{\begingroup \@sanitize@url \@href}%
\providecommand \@href[1]{\@@startlink{#1}\@@href}%
\providecommand \@@href[1]{\endgroup#1\@@endlink}%
\providecommand \@sanitize@url [0]{\catcode `\\12\catcode `\$12\catcode
  `\&12\catcode `\#12\catcode `\^12\catcode `\_12\catcode `\%12\relax}%
\providecommand \@@startlink[1]{}%
\providecommand \@@endlink[0]{}%
\providecommand \url  [0]{\begingroup\@sanitize@url \@url }%
\providecommand \@url [1]{\endgroup\@href {#1}{\urlprefix }}%
\providecommand \urlprefix  [0]{URL }%
\providecommand \Eprint [0]{\href }%
\providecommand \doibase [0]{http://dx.doi.org/}%
\providecommand \selectlanguage [0]{\@gobble}%
\providecommand \bibinfo  [0]{\@secondoftwo}%
\providecommand \bibfield  [0]{\@secondoftwo}%
\providecommand \translation [1]{[#1]}%
\providecommand \BibitemOpen [0]{}%
\providecommand \bibitemStop [0]{}%
\providecommand \bibitemNoStop [0]{.\EOS\space}%
\providecommand \EOS [0]{\spacefactor3000\relax}%
\providecommand \BibitemShut  [1]{\csname bibitem#1\endcsname}%
\let\auto@bib@innerbib\@empty
\bibitem [{\citenamefont {{Binney}}\ and\ \citenamefont
  {{Tremaine}}(2008)}]{2008gady.book.....B}%
  \BibitemOpen
  \bibfield  {author} {\bibinfo {author} {\bibfnamefont {J.}~\bibnamefont
  {{Binney}}}\ and\ \bibinfo {author} {\bibfnamefont {S.}~\bibnamefont
  {{Tremaine}}},\ }\href@noop {} {\emph {\bibinfo {title} {{Galactic Dynamics:
  Second Edition}}}}\ (\bibinfo {year} {2008})\BibitemShut {NoStop}%
\bibitem [{\citenamefont {Rubin}\ \emph {et~al.}(1978)\citenamefont {Rubin},
  \citenamefont {Ford~Jr},\ and\ \citenamefont {Thonnard}}]{rubin1978extended}%
  \BibitemOpen
  \bibfield  {author} {\bibinfo {author} {\bibfnamefont {V.~C.}\ \bibnamefont
  {Rubin}}, \bibinfo {author} {\bibfnamefont {W.~K.}\ \bibnamefont {Ford~Jr}},
  \ and\ \bibinfo {author} {\bibfnamefont {N.}~\bibnamefont {Thonnard}},\
  }\href@noop {} {\bibfield  {journal} {\bibinfo  {journal} {The Astrophysical
  Journal}\ }\textbf {\bibinfo {volume} {225}},\ \bibinfo {pages} {L107}
  (\bibinfo {year} {1978})}\BibitemShut {NoStop}%
\bibitem [{\citenamefont {Sofue}\ and\ \citenamefont
  {Rubin}(2001)}]{sofue2001rotation}%
  \BibitemOpen
  \bibfield  {author} {\bibinfo {author} {\bibfnamefont {Y.}~\bibnamefont
  {Sofue}}\ and\ \bibinfo {author} {\bibfnamefont {V.}~\bibnamefont {Rubin}},\
  }\href@noop {} {\bibfield  {journal} {\bibinfo  {journal} {Annual Review of
  Astronomy and Astrophysics}\ }\textbf {\bibinfo {volume} {39}},\ \bibinfo
  {pages} {137} (\bibinfo {year} {2001})}\BibitemShut {NoStop}%
\bibitem [{\citenamefont {Strigari}(2013)}]{strigari2013galactic}%
  \BibitemOpen
  \bibfield  {author} {\bibinfo {author} {\bibfnamefont {L.~E.}\ \bibnamefont
  {Strigari}},\ }\href@noop {} {\bibfield  {journal} {\bibinfo  {journal}
  {Physics Reports}\ }\textbf {\bibinfo {volume} {531}},\ \bibinfo {pages} {1}
  (\bibinfo {year} {2013})}\BibitemShut {NoStop}%
\bibitem [{\citenamefont {Amendola}\ \emph {et~al.}(2018)\citenamefont
  {Amendola}, \citenamefont {Appleby}, \citenamefont {Avgoustidis},
  \citenamefont {Bacon}, \citenamefont {Baker}, \citenamefont {Baldi},
  \citenamefont {Bartolo}, \citenamefont {Blanchard}, \citenamefont {Bonvin},
  \citenamefont {Borgani} \emph {et~al.}}]{amendola2018cosmology}%
  \BibitemOpen
  \bibfield  {author} {\bibinfo {author} {\bibfnamefont {L.}~\bibnamefont
  {Amendola}}, \bibinfo {author} {\bibfnamefont {S.}~\bibnamefont {Appleby}},
  \bibinfo {author} {\bibfnamefont {A.}~\bibnamefont {Avgoustidis}}, \bibinfo
  {author} {\bibfnamefont {D.}~\bibnamefont {Bacon}}, \bibinfo {author}
  {\bibfnamefont {T.}~\bibnamefont {Baker}}, \bibinfo {author} {\bibfnamefont
  {M.}~\bibnamefont {Baldi}}, \bibinfo {author} {\bibfnamefont
  {N.}~\bibnamefont {Bartolo}}, \bibinfo {author} {\bibfnamefont
  {A.}~\bibnamefont {Blanchard}}, \bibinfo {author} {\bibfnamefont
  {C.}~\bibnamefont {Bonvin}}, \bibinfo {author} {\bibfnamefont
  {S.}~\bibnamefont {Borgani}},  \emph {et~al.},\ }\href@noop {} {\bibfield
  {journal} {\bibinfo  {journal} {Living reviews in relativity}\ }\textbf
  {\bibinfo {volume} {21}},\ \bibinfo {pages} {1} (\bibinfo {year}
  {2018})}\BibitemShut {NoStop}%
\bibitem [{\citenamefont {Ciotti}(2022)}]{Ciotti:2022inn}%
  \BibitemOpen
  \bibfield  {author} {\bibinfo {author} {\bibfnamefont {L.}~\bibnamefont
  {Ciotti}},\ }\href {\doibase 10.3847/1538-4357/ac82b3} {\bibfield  {journal}
  {\bibinfo  {journal} {Astrophys. J.}\ }\textbf {\bibinfo {volume} {936}},\
  \bibinfo {pages} {180} (\bibinfo {year} {2022})},\ \Eprint
  {http://arxiv.org/abs/2207.09736} {arXiv:2207.09736 [astro-ph.GA]}
  \BibitemShut {NoStop}%
\bibitem [{\citenamefont {{Ostriker}}\ and\ \citenamefont
  {{Peebles}}(1973)}]{ostriker}%
  \BibitemOpen
  \bibfield  {author} {\bibinfo {author} {\bibfnamefont {J.~P.}\ \bibnamefont
  {{Ostriker}}}\ and\ \bibinfo {author} {\bibfnamefont {P.~J.~E.}\ \bibnamefont
  {{Peebles}}},\ }\href {\doibase 10.1086/152513} {\bibfield  {journal}
  {\bibinfo  {journal} {\apj}\ }\textbf {\bibinfo {volume} {186}},\ \bibinfo
  {pages} {467} (\bibinfo {year} {1973})}\BibitemShut {NoStop}%
\bibitem [{\citenamefont {Cooperstock}\ and\ \citenamefont
  {Tieu}(2007)}]{Cooperstock:2006dt}%
  \BibitemOpen
  \bibfield  {author} {\bibinfo {author} {\bibfnamefont {F.~I.}\ \bibnamefont
  {Cooperstock}}\ and\ \bibinfo {author} {\bibfnamefont {S.}~\bibnamefont
  {Tieu}},\ }\href {\doibase 10.1142/S0217751X0703666X} {\bibfield  {journal}
  {\bibinfo  {journal} {Int. J. Mod. Phys. A}\ }\textbf {\bibinfo {volume}
  {22}},\ \bibinfo {pages} {2293} (\bibinfo {year} {2007})},\ \Eprint
  {http://arxiv.org/abs/astro-ph/0610370} {arXiv:astro-ph/0610370} \BibitemShut
  {NoStop}%
\bibitem [{\citenamefont {Balasin}\ and\ \citenamefont
  {Grumiller}(2008)}]{Balasin:2006cg}%
  \BibitemOpen
  \bibfield  {author} {\bibinfo {author} {\bibfnamefont {H.}~\bibnamefont
  {Balasin}}\ and\ \bibinfo {author} {\bibfnamefont {D.}~\bibnamefont
  {Grumiller}},\ }\href {\doibase 10.1142/S0218271808012140} {\bibfield
  {journal} {\bibinfo  {journal} {Int. J. Mod. Phys. D}\ }\textbf {\bibinfo
  {volume} {17}},\ \bibinfo {pages} {475} (\bibinfo {year} {2008})},\ \Eprint
  {http://arxiv.org/abs/astro-ph/0602519} {arXiv:astro-ph/0602519} \BibitemShut
  {NoStop}%
\bibitem [{\citenamefont {Cross}(2006)}]{Cross:2006rx}%
  \BibitemOpen
  \bibfield  {author} {\bibinfo {author} {\bibfnamefont {D.~J.}\ \bibnamefont
  {Cross}},\ }\href@noop {} {\  (\bibinfo {year} {2006})},\ \Eprint
  {http://arxiv.org/abs/astro-ph/0601191} {arXiv:astro-ph/0601191} \BibitemShut
  {NoStop}%
\bibitem [{\citenamefont {Menzies}\ and\ \citenamefont
  {Mathews}(2007)}]{Menzies:2007dm}%
  \BibitemOpen
  \bibfield  {author} {\bibinfo {author} {\bibfnamefont {D.}~\bibnamefont
  {Menzies}}\ and\ \bibinfo {author} {\bibfnamefont {G.~J.}\ \bibnamefont
  {Mathews}},\ }\href@noop {} {\  (\bibinfo {year} {2007})},\ \Eprint
  {http://arxiv.org/abs/astro-ph/0701019} {arXiv:astro-ph/0701019} \BibitemShut
  {NoStop}%
\bibitem [{\citenamefont {Astesiano}\ \emph {et~al.}(2022)\citenamefont
  {Astesiano}, \citenamefont {Cacciatori}, \citenamefont {Gorini},\ and\
  \citenamefont {Re}}]{Astesiano:2021ren}%
  \BibitemOpen
  \bibfield  {author} {\bibinfo {author} {\bibfnamefont {D.}~\bibnamefont
  {Astesiano}}, \bibinfo {author} {\bibfnamefont {S.~L.}\ \bibnamefont
  {Cacciatori}}, \bibinfo {author} {\bibfnamefont {V.}~\bibnamefont {Gorini}},
  \ and\ \bibinfo {author} {\bibfnamefont {F.}~\bibnamefont {Re}},\ }\href
  {\doibase 10.1140/epjc/s10052-022-10506-7} {\bibfield  {journal} {\bibinfo
  {journal} {Eur. Phys. J. C}\ }\textbf {\bibinfo {volume} {82}},\ \bibinfo
  {pages} {554} (\bibinfo {year} {2022})},\ \Eprint
  {http://arxiv.org/abs/2106.12818} {arXiv:2106.12818 [gr-qc]} \BibitemShut
  {NoStop}%
\bibitem [{\citenamefont {Crosta}\ \emph {et~al.}(2020)\citenamefont {Crosta},
  \citenamefont {Giammaria}, \citenamefont {Lattanzi},\ and\ \citenamefont
  {Poggio}}]{crosta2020testing}%
  \BibitemOpen
  \bibfield  {author} {\bibinfo {author} {\bibfnamefont {M.}~\bibnamefont
  {Crosta}}, \bibinfo {author} {\bibfnamefont {M.}~\bibnamefont {Giammaria}},
  \bibinfo {author} {\bibfnamefont {M.~G.}\ \bibnamefont {Lattanzi}}, \ and\
  \bibinfo {author} {\bibfnamefont {E.}~\bibnamefont {Poggio}},\ }\href@noop {}
  {\bibfield  {journal} {\bibinfo  {journal} {Monthly Notices of the Royal
  Astronomical Society}\ }\textbf {\bibinfo {volume} {496}},\ \bibinfo {pages}
  {2107} (\bibinfo {year} {2020})}\BibitemShut {NoStop}%
\bibitem [{\citenamefont {{Gaia Collaboration}}(2016)}]{GAIA1}%
  \BibitemOpen
  \bibfield  {author} {\bibinfo {author} {\bibnamefont {{Gaia
  Collaboration}}},\ }\href@noop {} {\bibfield  {journal} {\bibinfo  {journal}
  {Astronomy \& Astrophysics}\ }\textbf {\bibinfo {volume} {595}},\ \bibinfo
  {pages} {A1} (\bibinfo {year} {2016})}\BibitemShut {NoStop}%
\bibitem [{\citenamefont {{Gaia Collaboration}}(2018)}]{GAIA2}%
  \BibitemOpen
  \bibfield  {author} {\bibinfo {author} {\bibnamefont {{Gaia
  Collaboration}}},\ }\href@noop {} {\bibfield  {journal} {\bibinfo  {journal}
  {Astronomy \& Astrophysics}\ }\textbf {\bibinfo {volume} {616}},\ \bibinfo
  {pages} {A1} (\bibinfo {year} {2018})}\BibitemShut {NoStop}%
\bibitem [{\citenamefont {Ramos-Caro}\ \emph {et~al.}(2012)\citenamefont
  {Ramos-Caro}, \citenamefont {Agon},\ and\ \citenamefont
  {Pedraza}}]{Ramos-Caro:2012ren}%
  \BibitemOpen
  \bibfield  {author} {\bibinfo {author} {\bibfnamefont {J.}~\bibnamefont
  {Ramos-Caro}}, \bibinfo {author} {\bibfnamefont {C.~A.}\ \bibnamefont
  {Agon}}, \ and\ \bibinfo {author} {\bibfnamefont {J.~F.}\ \bibnamefont
  {Pedraza}},\ }\href {\doibase 10.1103/PhysRevD.86.043008} {\bibfield
  {journal} {\bibinfo  {journal} {Phys. Rev. D}\ }\textbf {\bibinfo {volume}
  {86}},\ \bibinfo {pages} {043008} (\bibinfo {year} {2012})},\ \Eprint
  {http://arxiv.org/abs/1206.5804} {arXiv:1206.5804 [gr-qc]} \BibitemShut
  {NoStop}%
\bibitem [{\citenamefont {Ludwig}(2021)}]{ludwig2021galactic}%
  \BibitemOpen
  \bibfield  {author} {\bibinfo {author} {\bibfnamefont {G.}~\bibnamefont
  {Ludwig}},\ }\href@noop {} {\bibfield  {journal} {\bibinfo  {journal} {The
  European Physical Journal C}\ }\textbf {\bibinfo {volume} {81}},\ \bibinfo
  {pages} {1} (\bibinfo {year} {2021})}\BibitemShut {NoStop}%
\bibitem [{\citenamefont {Ciufolini}\ and\ \citenamefont
  {Wheeler}(1995)}]{ciufolini1995gravitation}%
  \BibitemOpen
  \bibfield  {author} {\bibinfo {author} {\bibfnamefont {I.}~\bibnamefont
  {Ciufolini}}\ and\ \bibinfo {author} {\bibfnamefont {J.~A.}\ \bibnamefont
  {Wheeler}},\ }\href@noop {} {\emph {\bibinfo {title} {Gravitation and
  inertia}}},\ Vol.\ \bibinfo {volume} {101}\ (\bibinfo  {publisher} {Princeton
  university press},\ \bibinfo {year} {1995})\BibitemShut {NoStop}%
\bibitem [{\citenamefont {Mashhoon}\ \emph {et~al.}(2001)\citenamefont
  {Mashhoon}, \citenamefont {Gronwald},\ and\ \citenamefont
  {Lichtenegger}}]{Mashhoon:2001ir}%
  \BibitemOpen
  \bibfield  {author} {\bibinfo {author} {\bibfnamefont {B.}~\bibnamefont
  {Mashhoon}}, \bibinfo {author} {\bibfnamefont {F.}~\bibnamefont {Gronwald}},
  \ and\ \bibinfo {author} {\bibfnamefont {H.~I.~M.}\ \bibnamefont
  {Lichtenegger}},\ }in\ \href@noop {} {\emph {\bibinfo {booktitle} {Gyros,
  Clocks, Interferometers...: Testing Relativistic Graviy in Space}}}\
  (\bibinfo  {publisher} {Springer, Berlin, Heidelberg},\ \bibinfo {address}
  {Berlin, Heidelberg},\ \bibinfo {year} {2001})\ pp.\ \bibinfo {pages}
  {83--108}\BibitemShut {NoStop}%
\bibitem [{\citenamefont {Ruggiero}\ and\ \citenamefont
  {Tartaglia}(2002)}]{Ruggiero:2002hz}%
  \BibitemOpen
  \bibfield  {author} {\bibinfo {author} {\bibfnamefont {M.~L.}\ \bibnamefont
  {Ruggiero}}\ and\ \bibinfo {author} {\bibfnamefont {A.}~\bibnamefont
  {Tartaglia}},\ }\href@noop {} {\bibfield  {journal} {\bibinfo  {journal}
  {Nuovo Cim.}\ }\textbf {\bibinfo {volume} {B117}},\ \bibinfo {pages} {743}
  (\bibinfo {year} {2002})},\ \Eprint {http://arxiv.org/abs/gr-qc/0207065}
  {arXiv:gr-qc/0207065 [gr-qc]} \BibitemShut {NoStop}%
\bibitem [{\citenamefont {Mashhoon}(2003)}]{Mashhoon:2003ax}%
  \BibitemOpen
  \bibfield  {author} {\bibinfo {author} {\bibfnamefont {B.}~\bibnamefont
  {Mashhoon}},\ }\href@noop {} {\  (\bibinfo {year} {2003})},\ \Eprint
  {http://arxiv.org/abs/gr-qc/0311030} {arXiv:gr-qc/0311030 [gr-qc]}
  \BibitemShut {NoStop}%
\bibitem [{\citenamefont {Costa}\ and\ \citenamefont
  {Natario}(2014)}]{Costa:2012cw}%
  \BibitemOpen
  \bibfield  {author} {\bibinfo {author} {\bibfnamefont {L.~F.~O.}\
  \bibnamefont {Costa}}\ and\ \bibinfo {author} {\bibfnamefont
  {J.}~\bibnamefont {Natario}},\ }\href {\doibase 10.1007/s10714-014-1792-1}
  {\bibfield  {journal} {\bibinfo  {journal} {Gen. Rel. Grav.}\ }\textbf
  {\bibinfo {volume} {46}},\ \bibinfo {pages} {1792} (\bibinfo {year}
  {2014})},\ \Eprint {http://arxiv.org/abs/1207.0465} {arXiv:1207.0465 [gr-qc]}
  \BibitemShut {NoStop}%
\bibitem [{\citenamefont {Astesiano}\ and\ \citenamefont
  {Ruggiero}(2022)}]{Astesiano:2022ozl}%
  \BibitemOpen
  \bibfield  {author} {\bibinfo {author} {\bibfnamefont {D.}~\bibnamefont
  {Astesiano}}\ and\ \bibinfo {author} {\bibfnamefont {M.~L.}\ \bibnamefont
  {Ruggiero}},\ }\href {\doibase 10.1103/PhysRevD.106.044061} {\bibfield
  {journal} {\bibinfo  {journal} {Phys. Rev. D}\ }\textbf {\bibinfo {volume}
  {106}},\ \bibinfo {pages} {044061} (\bibinfo {year} {2022})},\ \Eprint
  {http://arxiv.org/abs/2205.03091} {arXiv:2205.03091 [gr-qc]} \BibitemShut
  {NoStop}%
\bibitem [{\citenamefont {Collier}(2019)}]{10.1093/mnras/stz3625}%
  \BibitemOpen
  \bibfield  {author} {\bibinfo {author} {\bibfnamefont {A.}~\bibnamefont
  {Collier}},\ }\href {\doibase 10.1093/mnras/stz3625} {\bibfield  {journal}
  {\bibinfo  {journal} {MNRAS}\ }\textbf {\bibinfo {volume} {492}},\ \bibinfo
  {pages} {2241} (\bibinfo {year} {2019})}\BibitemShut {NoStop}%
\bibitem [{\citenamefont {McMillan}\ \emph {et~al.}(2022)\citenamefont
  {McMillan}, \citenamefont {Petersson}, \citenamefont {Tepper-Garcia},
  \citenamefont {Bland-Hawthorn}, \citenamefont {Antoja}, \citenamefont
  {Chemin}, \citenamefont {Figueras}, \citenamefont {Khanna}, \citenamefont
  {Kordopatis}, \citenamefont {Ramos}, \citenamefont {Romero-G{\'o}mez},\ and\
  \citenamefont {Seabroke}}]{10.1093/mnras/stac2571}%
  \BibitemOpen
  \bibfield  {author} {\bibinfo {author} {\bibfnamefont {P.~J.}\ \bibnamefont
  {McMillan}}, \bibinfo {author} {\bibfnamefont {J.}~\bibnamefont {Petersson}},
  \bibinfo {author} {\bibfnamefont {T.}~\bibnamefont {Tepper-Garcia}}, \bibinfo
  {author} {\bibfnamefont {J.}~\bibnamefont {Bland-Hawthorn}}, \bibinfo
  {author} {\bibfnamefont {T.}~\bibnamefont {Antoja}}, \bibinfo {author}
  {\bibfnamefont {L.}~\bibnamefont {Chemin}}, \bibinfo {author} {\bibfnamefont
  {F.}~\bibnamefont {Figueras}}, \bibinfo {author} {\bibfnamefont
  {S.}~\bibnamefont {Khanna}}, \bibinfo {author} {\bibfnamefont
  {G.}~\bibnamefont {Kordopatis}}, \bibinfo {author} {\bibfnamefont
  {P.}~\bibnamefont {Ramos}}, \bibinfo {author} {\bibfnamefont
  {M.}~\bibnamefont {Romero-G{\'o}mez}}, \ and\ \bibinfo {author}
  {\bibfnamefont {G.}~\bibnamefont {Seabroke}},\ }\href {\doibase
  10.1093/mnras/stac2571} {\bibfield  {journal} {\bibinfo  {journal} {MNRAS}\
  }\textbf {\bibinfo {volume} {516}},\ \bibinfo {pages} {4988} (\bibinfo {year}
  {2022})}\BibitemShut {NoStop}%
\bibitem [{\citenamefont {Ruggiero}\ \emph {et~al.}(2022)\citenamefont
  {Ruggiero}, \citenamefont {Ortolan},\ and\ \citenamefont
  {Speake}}]{Ruggiero:2021lpf}%
  \BibitemOpen
  \bibfield  {author} {\bibinfo {author} {\bibfnamefont {M.~L.}\ \bibnamefont
  {Ruggiero}}, \bibinfo {author} {\bibfnamefont {A.}~\bibnamefont {Ortolan}}, \
  and\ \bibinfo {author} {\bibfnamefont {C.~C.}\ \bibnamefont {Speake}},\
  }\href {\doibase 10.1088/1361-6382/ac9949} {\bibfield  {journal} {\bibinfo
  {journal} {Class. Quant. Grav.}\ }\textbf {\bibinfo {volume} {39}},\ \bibinfo
  {pages} {225015} (\bibinfo {year} {2022})},\ \Eprint
  {http://arxiv.org/abs/2112.08290} {arXiv:2112.08290 [gr-qc]} \BibitemShut
  {NoStop}%
\bibitem [{\citenamefont {Mars}\ and\ \citenamefont
  {Senovilla}(1998)}]{Mars:1998qd}%
  \BibitemOpen
  \bibfield  {author} {\bibinfo {author} {\bibfnamefont {M.}~\bibnamefont
  {Mars}}\ and\ \bibinfo {author} {\bibfnamefont {J.~M.~M.}\ \bibnamefont
  {Senovilla}},\ }\href {\doibase 10.1142/S0217732398001583} {\bibfield
  {journal} {\bibinfo  {journal} {Mod. Phys. Lett. A}\ }\textbf {\bibinfo
  {volume} {13}},\ \bibinfo {pages} {1509} (\bibinfo {year} {1998})},\ \Eprint
  {http://arxiv.org/abs/gr-qc/9806094} {arXiv:gr-qc/9806094} \BibitemShut
  {NoStop}%
\bibitem [{\citenamefont {Astesiano}(2022)}]{Astesiano:2022gph}%
  \BibitemOpen
  \bibfield  {author} {\bibinfo {author} {\bibfnamefont {D.}~\bibnamefont
  {Astesiano}},\ }\href {\doibase 10.1007/s10714-022-02947-y} {\bibfield
  {journal} {\bibinfo  {journal} {Gen. Rel. Grav.}\ }\textbf {\bibinfo {volume}
  {54}},\ \bibinfo {pages} {63} (\bibinfo {year} {2022})},\ \Eprint
  {http://arxiv.org/abs/2201.03959} {arXiv:2201.03959 [gr-qc]} \BibitemShut
  {NoStop}%
\end{thebibliography}%

\end{document}